\documentclass[twocolumn]{jpsj3} 

\def\runauthor{T. \textsc{Sasaki} {\it et al.}}

\title{Suppression of Superconductivity by Nonmagnetic Disorder in Organic Superconductor $\kappa$-(BEDT-TTF)$_{2}$Cu(NCS)$_{2}$}

\author{
Takahiko \textsc{Sasaki}$^{1,2}$\thanks{E-mail address: takahiko@imr.tohoku.ac.jp}, Hajime \textsc{Oizumi}$^{1}$\thanks{Present address: Suzuki Motor Corporation}, Yuki \textsc{Honda}$^{1}$\thanks{Present address: Harmonic Drive Systems inc.}, Naoki \textsc{Yoneyama}$^{1,2}$\thanks{Present address: Interdisciplinary Graduate School of Medical and Engineering, University of Yamanashi, Japan}, and Norio \textsc{Kobayashi}$^{1}$
}

\inst{%
$^{1}$Institute for Materials Research, Tohoku University, Sendai 980-8577, Japan\\

$^{2}$Japan Science and Technology Agency, CREST, Tokyo 102-0075, Japan\\
}
\recdate{\today}

\abst{
The suppression of superconductivity by nonmagnetic disorder is investigated systematically in the organic superconductor $\kappa$-(BEDT-TTF)$_2$Cu(NCS)$_2$.  
We introduce a nonmagnetic disorder arising from molecule substitution in part with deuterated BEDT-TTF or BMDT-TTF for BEDT-TTF molecules and molecular defects introduced by X-ray irradiation.  
A quantitative evaluation of the scattering time $\tau_{\rm dHvA}$ is carried out by de Haas-van Alphen (dHvA) effect measurement.
A large reduction in $T_{\rm c}$ with a linear dependence on $1/\tau_{\rm dHvA}$ is found in the small-disorder region below $1/\tau_{\rm dHvA} \simeq$ 1 $\times$ 10$^{12}$ s$^{-1}$ in both the BMDT-TTF molecule-substituted and X-ray-irradiated samples. 
The observed linear relation between $T_{\rm c}$ and $1/\tau_{\rm dHvA}$ is in agreement with the Abrikosov-Gorkov (AG) formula, at least in the small-disorder region.  
This observation is reasonably consistent with the unconventional superconductivity proposed thus far for the present organic superconductor.  
A deviation from the AG formula, however, is observed in the large-disorder region above $1/\tau_{\rm dHvA} \simeq$ 1 $\times$ 10$^{12}$ s$^{-1}$, which reproduces the previous transport study (J. G. Analytis {\it et al.}: Phys. Rev. Lett. {\bf 96} (2006) 177002). 
We present some interpretations of this deviation from the viewpoints of superconductivity and the inherent difficulties in the evaluation of scattering time. 
}

\kword{
organic superconductor, disorder effect, $\kappa$-(BEDT-TTF)$_{2}$Cu(NCS)$_{2}$, de Haas-van Alphen oscillations, scattering time, X-ray irradiation
}

\begin{document}
\maketitle

\section{Introduction} 

Superconductivity in organic charge transfer salts, $\kappa$-(BEDT-TTF)$_2X$, where BEDT-TTF (or ET in short) denotes bis(ethylenedithio)tetrathiafulvalene, has been investigated extensively \cite{kanoda1,lang}.  
Its remarkable feature is that the native quarter-filled band is modified to the effective half-filled band by a strong dimer structure consisting of two BEDT-TTF molecules.
Thus, this family of organic conductors has been considered as a typical bandwidth-controlled Mott transition system with strongly correlated electrons \cite{kanoda2,kino}.
In the $\kappa$-(BEDT-TTF)$_2X$ system, the bandwidth can be controlled by applying physical pressure and a slight chemical substitution of the molecules, which can change the conduction bandwidth $W$ with respect to the effective Coulomb repulsion $U$ between two electrons on a dimer.  
A first-order electronic phase transition line divides the phase diagram into the superconducting and antiferromagnetic (AF) Mott insulator phases \cite{kanoda2,kagawa1,sasaki4}.  
The superconductivity near an AF Mott insulator has been theoretically expected to possess an unconventional $d$-wave order parameter \cite{kondo,kuroki}.  
Experimental results on unconventional superconductivity, however, have not been elucidated yet \cite{lang}. 

To clarify the pair-breaking mechanism, the investigation of the disorder effect on superconductivity is important because it can give information on the pairing symmetry \cite{anderson,maple,millis,radtke}.  
There have been several attempts to introduce disorder into organic superconductors, which arises from molecular defects and impurities by electron, proton, and X-ray irradiation \cite{analytis,dolanski,sasaki5} and anion $X$ or donor molecule substitution \cite{naito,yone1,yone3}, respectively.
In addition, the conformational disorder of terminal ethylene groups in the BEDT-TTF molecule has been investigated from the viewpoint of the cooling speed dependence of electronic properties at low temperatures \cite{stalcup,su,yone2}.
In the case of organic superconductors, however, there are characteristic difficulties in the experimental investigations of the disorder effect.  
Superconductivity is very sensitive to the applied pressure and strain, probably owing to the softness of the molecules and molecular lattices.  
For example, the hydrostatic pressure dependence of $T_{\rm c}$ becomes approximately -3 K/kbar in $\kappa$-(BEDT-TTF)$_{2}$Cu(NCS)$_{2}$ \cite{schirber}, which is roughly two orders of magnitude larger than that of inorganic metals, for example, -0.035 K/kbar for Pb \cite{Pb}.  
In addition, it has been revealed from Ehrenfest analysis for thermal expansion measurements \cite{mueller2002} that the uniaxial stress/strain effect on $T_{\rm c}$ is quite large and anisotropic for the crystal axes.
A large negative uniaxial pressure component of approximately -6.2 K/kbar is found for the interplane direction, while intraplane components indicate positive values of +3.4 and +0.14 K/kbar in $\kappa$-(BEDT-TTF)$_{2}$Cu(NCS)$_{2}$.  
Such a large structural effect on superconductivity makes the systematic study of the disorder effect difficult in general.

In this paper, we present a study of the disorder effect on the superconductivity of the organic superconductor $\kappa$-(BEDT-TTF)$_{2}$Cu(NCS)$_{2}$, which was synthesized as the first 10 K class organic superconductor.\cite{urayama}  
Three different types of disorder are introduced by the partial substitution of BEDT-TTF molecules with deuterated BEDT-TTF or BMDT-TTF molecules, where BMDT-TTF (or MT) denotes bis(methylenedithio)tetrathiafulvalene, and molecular defects induced by X-ray irradiation into the crystals.  
To discuss the relation between $T_{\rm c}$ and disorder systematically, we carried out de Haas-van Alphen (dHvA) measurement for the quantitative evaluation of the disorder level using the scattering time, which could be obtained more reliably without the complex assumptions needed in a transport study. 
In addition, $T_{\rm c}$ is determined thermodynamically by measuring the magnetic susceptibility. 

From the results, a large reduction in $T_{\rm c}$ with a linear dependence on $1/\tau_{\rm dHvA}$, evaluated on the basis of the dHvA effect, is found in the small-disorder region below $1/\tau_{\rm dHvA} \simeq$ 1 $\times$ 10$^{12}$ s$^{-1}$ in both the BMDT-TTF-molecule-substituted and X-ray-irradiated samples. 
The observed linear relation between $T_{\rm c}$ and $1/\tau_{\rm dHvA}$ is in agreement with the Abrikosov-Gorkov (AG) formula \cite{AG} at least in the small-disorder region.  
A possible explanation for this observation may be the nonmagnetic disorder effect on $T_{\rm c}$ for the unconventional superconductivity discussed so far in this organic superconductor. \cite{powell}
A deviation from the AG formula, however, is observed in the large-disorder region, which reproduces the results of a previous transport study \cite{analytis}. 
We present some interpretations of this deviation from the viewpoints of superconductivity and the inherent difficulties in the evaluation of the scattering time. 

\section{Experiment} 

Single crystals of $\kappa$-($h$-BEDT-TTF)$_{2}$Cu(NCS)$_{2}$, $\kappa$-[($h$-BEDT-TTF)$_{1-x}$($d$-BEDT-TTF)$_{x}$]$_{2}$Cu(NCS)$_{2}$, and $\kappa$-[($h$-BEDT-TTF)$_{1-x}$($h$-BMDT-TTF)$_{x}$]$_{2}$Cu(NCS)$_{2}$ were grown by an electrochemical oxidation method, where $h$ and $d$-BEDT-TTF denote the hydrogenated and deuterated BEDT-TTF molecules, respectively. 
Crystals with BMDT-TTF molecule substitution could be grown up to a substitution ratio of $x =$ 0.15,\cite{yone3,MT-lattice} while crystals with partial substitution of the deuterated BEDT-TTF molecule were obtained at $x =$ 0 -- 1.
The substitution ratios $x$ of $d$-BEDT-TTF and $h$-BMDT-TTF for $h$-BEDT-TTF were examined by analyzing the molecular vibration of the terminal ethylene group of $h$-BEDT-TTF in the infrared reflectance spectra.
The magnitude of the vibration modes for the terminal ethylenes should change linearly with the number of terminal ethylenes \cite{sasaki2009stam}.  
The actual substitution ratio was confirmed by this method and the measured ratio $x$ was found to be almost the same as the nominal value in the crystal growth.
For the X-ray irradiation experiments, a crystal of $\kappa$-($h$-BEDT-TTF)$_{2}$Cu(NCS)$_{2}$ was irradiated at 300 K using a nonfiltered tungsten target at 40 kV and 20 mA.  
The dose rate under the present irradiation conditions was expected to be about 0.5 MGy/h following a comparison with a previous report by Analytis {\it et al.}\cite{analytis}
The magnetic susceptibility measurements using a superconducting quantum interference device (SQUID) magnetometer at 5 T showed no indication of the production of magnetic impurities by the molecule substitution and X-ray irradiation. \cite{x-ray_magnetic}  

The superconducting transition temperature $T_{\rm c}$ of the samples was determined from the magnetic susceptibility at 3 Oe.  
The definition of $T_{\rm c}$ is given in the next section in comparison to that used in the resistivity measurements.
Magnetic torque measurements were performed to detect dHvA oscillations using a precision capacitance torquemeter \cite{sasaki1998,sasaki2003}. 
The torquemeter with samples was directly immersed in liquid $^{3}$He or dense $^{3}$He gas in a refrigerator, which was combined with a 15 T superconducting magnet at the High Magnetic Field Laboratory for Superconducting Materials (HFLSM), IMR, Tohoku University. 
In all the measurements of the magnetic susceptibility and dHvA effect, the samples were cooled slowly from room temperature to 4.2 K in approximately 24 h in order to obtain the same sample conditions and exclude any possible disorder effect on the conformational order of the terminal ethylene groups of BEDT-TTF molecules \cite{stalcup,su,yone2,mueller2002,yone2005}.


\section{Results and Discussion}
\subsection{Disorder effect on superconductivity} 

\begin{figure}
\includegraphics[width=1\linewidth]{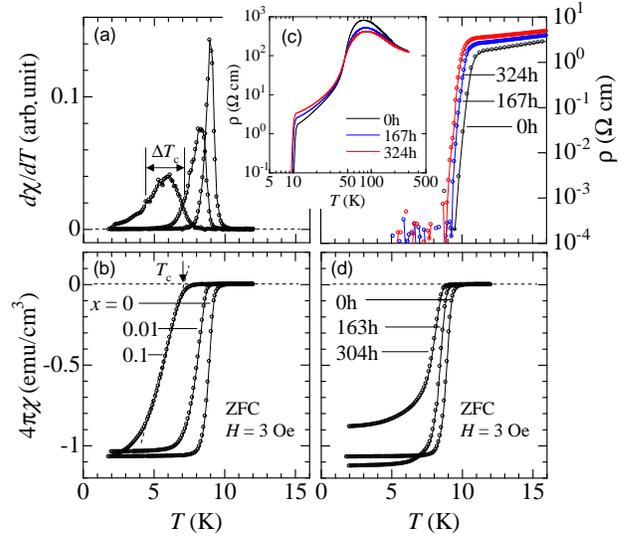}
\caption{(Color online) Temperature dependences of (b) the magnetic susceptibility $\chi$ and (a) the temperature derivative of $\chi$ in $\kappa$-[($h$-BEDT-TTF)$_{1-x}$($h$-BMDT-TTF)$_{x}$]$_{2}$Cu(NCS)$_{2}$, and (d) $\chi$ in the X-ray irradiated $\kappa$-($h$-BEDT-TTF)$_{2}$Cu(NCS)$_{2}$.  The measurements were performed at $H =$ 3 Oe perpendicular to the $b$-$c$ plane under the zero-field-cooling condition. The definitions of $T_{\rm c}$ and the width ${\Delta}T_{\rm c}$ are given in the text. (c) Temperature dependence of interlayer resistivity  of X-ray-irradiated $\kappa$-($h$-BEDT-TTF)$_{2}$Cu(NCS)$_{2}$. Note that the samples in (c) and (d) are different samples irradiated under almost the same conditions.}
\end{figure}

Figures 1(b) and 1(d) respectively show the temperature dependences of the magnetic susceptibilities $4\pi\chi$ of $\kappa$-[($h$-BEDT-TTF)$_{1-x}$($h$-BMDT-TTF)$_{x}$]$_{2}$Cu(NCS)$_{2}$ and the X-ray-irradiated $\kappa$-($h$-BEDT-TTF)$_{2}$Cu(NCS)$_{2}$.
The demagnetization effect is corrected by using an ellipsoidal approximation for the planar shape of the samples.
The ambiguity of the shape and size of the samples may lead to an error (approximately $\pm 15 \%$) for the full superconducting diamagnetic volume $-1/4\pi$ \cite{chi}.
$T_{\rm c}$ is determined from the diamagnetic transition curve as the crossing point of the interpolation lines from the normal and superconducting regions.  

The $T_{\rm c}$ of the pristine samples determined here reproduces well the previous results in which $T_{\rm c}$ is obtained thermodynamically by specific heat \cite{Andraka,Graebner,Muller_C,Taylor_C}, thermal expansion \cite{Lang_TE}, and magnetization \cite{Lang_PRL92,Lang_PRB94} measurements.  
In the resistivity measurements, in general, the offset temperature to a zero resistivity seems to be close to the $T_{\rm c}$ obtained from the thermodynamic measurements, although the determination of the offset temperature depends on the actual measurement accuracy of the sufficiently low resistivity.  
Figure 1(c) demonstrates the resistive superconducting transition curves of the X-ray irradiated $\kappa$-($h$-BEDT-TTF)$_{2}$Cu(NCS)$_{2}$.  
The resistivity is measured along the interlayer direction.
The changes of the temperature dependences of the resistivity and $T_{\rm c}$ with the irradiation reproduce well the previous results \cite{analytis}.  
In comparison with the magnetization results in Fig. 1(d), the offset criterion of the resistivity by 0.1\% of the normal resistivity indicates an almost the same $T_{\rm c}$ determined from the present magnetization measurements and the previous thermodynamically obtained $T_{\rm c}$ of a pristine sample.  
Note, therefore, that $T_{\rm c}$ in the present paper is lower by approximately 1 K than that determined as the middle point of the resistive transition curve \cite{analytis}. 

The width of the superconducting transition is evaluated to be the half-width of the temperature derivative curve $d\chi/dT$, as shown in Fig. 1(a) for $\kappa$-[($h$-BEDT-TTF)$_{1-x}$($d$-BEDT-TTF)$_{x}$]$_{2}$Cu(NCS)$_{2}$.  
These conditions and definitions are the same as those for the measurements of $\kappa$-[($h$-BEDT-TTF)$_{1-x}$($d$-BEDT-TTF)$_{x}$]$_{2}$Cu(NCS)$_{2}$ and the X-ray-irradiated $\kappa$-($h$-BEDT-TTF)$_{2}$Cu(NCS)$_{2}$.

\begin{figure}
\includegraphics[width=1\linewidth]{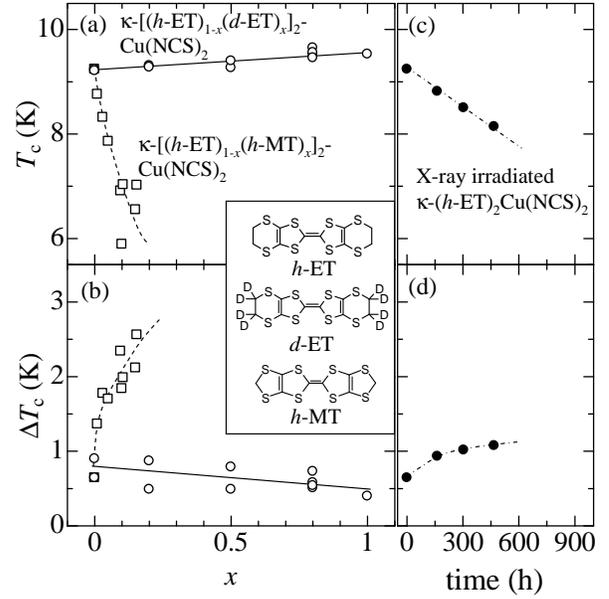}
\caption{Molecule substitution and X-ray irradiation time dependences of $T_{\rm c}$ and ${\Delta}T_{\rm c}$ for (a), (b) $\kappa$-[($h$-BEDT-TTF)$_{1-x}$($d$-BEDT-TTF)$_{x}$]$_{2}$Cu(NCS)$_{2}$ and $\kappa$-[($h$-BEDT-TTF)$_{1-x}$($h$-BMDT-TTF)$_{x}$]$_{2}$Cu(NCS)$_{2}$ and (c), (d) $\kappa$-($h$-BEDT-TTF)$_{2}$Cu(NCS)$_{2}$ irradiated with X-rays, respectively. The curves are guides for the eyes.  }
\end{figure}

The changes in $T_{\rm c}$ and ${\Delta}T_{\rm c}$ with the substitution ratio $x$ and irradiation time are summarized in Fig. 2.  
The substitution by BMDT-TTF molecules reduces $T_{\rm c}$ almost linearly with $x$ up to $\sim 0.15$.  
The transition width increases concurrently with a decrease in $T_{\rm c}$. 
The observed large suppression of the superconductivity may be caused by the disorder effect with the nonmagnetic BMDT-TTF molecule substitution. 
The increase in the transition width indicates that the BMDT-TTF molecule substitution also causes an inhomogeneity in the superconductivity.
On the other hand, the deuterated BEDT-TTF molecule substitution does not markedly affect superconductivity.  
In the full range of the substitution, the largest reduction in $T_{\rm c}$ by the disorder effect is expected to take place at $x =$ 0.5.  
The experimental results, however, show that $T_{\rm c}$ increases linearly from $x =$ 0 to 1 and ${\Delta}T_{\rm c}$ is almost constant.  
The increase in $T_{\rm c}$ results from a chemical pressure effect on the narrowing of the bandwidth $W$ with respect to $U$ on the dimer \cite{yone1}.  
Thus, we can conclude that the deuterated BEDT-TTF molecule substitution does not contribute at all to the suppression of the superconductivity.
This is also supported by the small change in the scattering time with $x$, which is obtained from the dHvA effect discussed in the next section.

The sample irradiated with X-rays shows a decrease in $T_{\rm c}$ with an increase in the irradiation time. 
The reduction in $T_{\rm c}$ reproduces well the results reported by Analytis {\it et al} \cite{analytis}.
In addition, the present magnetic susceptibility measurements indicate the broadening of ${\Delta}T_{\rm c}$ in connection with the reduction in $T_{\rm c}$ by X-ray irradiation.
The behaviors of $T_{\rm c}$ and ${\Delta}T_{\rm c}$ are similar to those in the samples with the BMDT-TTF substitution.  
This suggests that X-ray irradiation induces a disorder effect for superconductivity owing to the random potential modulation caused by defects in molecules.  
Considering the changes in $T_{\rm c}$ and ${\Delta}T_{\rm c}$ by the X-ray irradiation in comparison with the BMDT-TTF molecule substitution, 500 h X-ray irradiation roughly corresponds to the $x =$ 0.05 -- 0.1 substitution of BMDT-TTF molecules.  

To evaluate the disorder effect on quasi-particle scattering in metals, the residual resistivity $\rho_{\rm 0}$ at low temperatures is a good physical quantity for estimating the scattering time $\tau_{\rm imp}$ by impurity scattering as $\rho_{\rm 0} = m^{*}/ne^{2}\tau_{\rm imp}$, where $m^{*}$ is the effective mass, $n$ is the number of the carriers, and $e$ is the elementary electric charge.  
The scattering time $\tau_{\rm imp}$ obtained from $\rho_{0}$ contains mostly the contribution of large-angle scattering by impurities.  
The reduction in the $T_{\rm c}$ of $\kappa$-(BEDT-TTF)$_{2}$Cu(NCS)$_{2}$ irradiated with X-rays has been explained by the AG formula with the quasi-particle scattering time obtained from the residual resistivity \cite{analytis}.  
There are, however, some disadvantages of using the residual resistivity to obtain the scattering time.  
First, the scattering time is obtained indirectly from the resistivity because it is necessary to determine $m^{*}$ and $n$.  
Second, it is experimentally difficult to accurately obtain the intralayer resistivity for small samples with a large anisotropy between intra- and interlayer resistivities \cite{singleton1,strack}.  
In addition, we need to assume the interlayer transfer integral $t_{\perp}$ in the case that the interlayer resistivity is used for obtaining $\tau_{\rm imp}$ \cite{powell,mckenzie1998}.  
One of the experimental methods of quantitatively obtaining the scattering time is the measurement of the dHvA effect \cite{shoenberg}.  
The scattering time can be evaluated directly from dHvA oscillations in addition to the effective mass and the cross-sectional area of the Fermi surface.  
The scattering time $\tau_{\rm dHvA}$ obtained from the dHvA effect also includes contributions of small-angle scattering in addition to that of large-angle scattering such as elastic impurity scattering. 
We should consider different contributions to $\tau_{\rm dHvA}$ from $\tau_{\rm imp}$ for the quantitative consideration of physical properties such as $T_{\rm c}$.
The dHvA effect of the molecule-substituted and X-ray-irradiated $\kappa$-(BEDT-TTF)$_{2}$Cu(NCS)$_{2}$ is presented in the next section.

\subsection{Disorder effect on de Haas-van Alphen oscillations}

\begin{figure}
\includegraphics[width=1\linewidth]{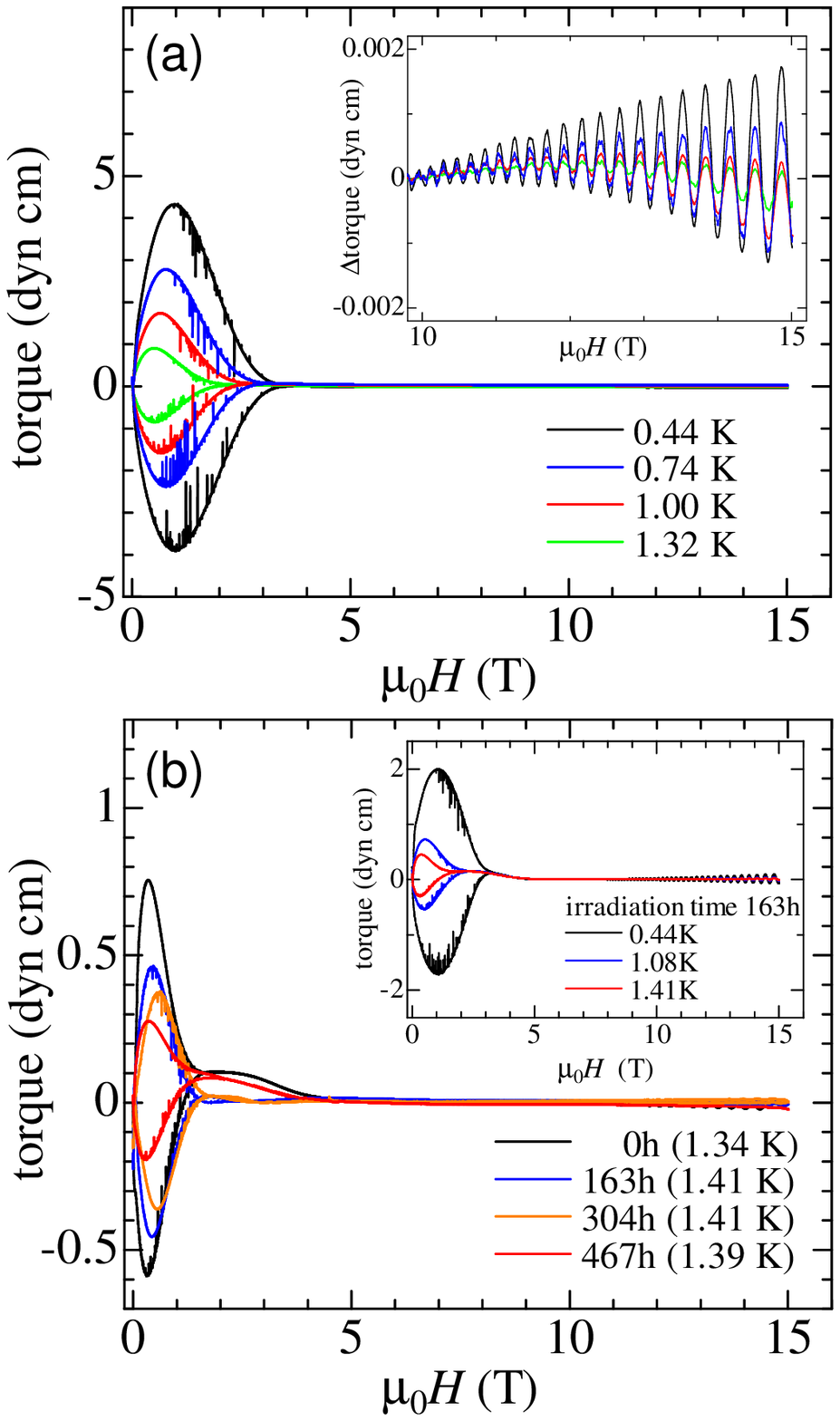}
\caption{(Color online) (a) Temperature dependence of magnetic torque curves of $\kappa$-[($h$-BEDT-TTF)$_{0.97}$($h$-BMDT-TTF)$_{0.03}$]$_{2}$Cu(NCS)$_{2}$.  The magnetic fields are applied perpendicular to the $b$-$c$ plane. The inset indicates the torque dHvA oscillations after subtracting the monotonic background torque curves. (b) Magnetic torque curves of $\kappa$-($h$-BEDT-TTF)$_{2}$Cu(NCS)$_{2}$ irradiated with X-rays with increasing irradiation dose. The temperature indicated in parentheses is the temperature for each torque measurements. A slightly different sample setting of the torquemeter and a slight difference in the magnetic field direction in each measurement after the irradiation at room temperature sensitively affect the amplitude and structure of torque curves, especially in the superconducting region below approximately 5 T. The inset of the figure shows the temperature dependence of the magnetic torque curves after 163 h of X-ray irradiation. }
\end{figure}

Figures 3(a) and 3(b) show the magnetic torque curves and dHvA oscillations of the BMDT-TTF-molecule-substituted sample with $x =$ 0.03 at 0.44 -- 1.32 K and of the X-ray-irradiated sample with different irradiation doses, respectively.  
The magnetic field is applied perpendicular to the conductive $b$-$c$ plane. 
At low magnetic fields, a large hysteresis in the torque curves appears for the up and down sweeps of the magnetic field because of the magnetic irreversibility owing to the vortex solid states of the type-II superconductor \cite{nishizaki}.  
The sharp spike structures observed inside the hysteresis loop result from the flux jumps appearing in the nonequilibrium vortex state upon sweeping the magnetic field \cite{sasaki1998,mola,konoike}.  
In the X-ray-irradiated sample, as shown in Fig. 3(b), the torque curves in the superconducting region show a measurement-to-measurement variation although the same sample is measured with increasing irradiation dose.  
The reason for the variation may be the fact that the sample is remounted onto the torquemeter after the sample is irradiated with X-rays at room temperature. 
A small difference in the setting condition between the sample direction and the magnetic field direction nearly perpendicular to the $b$-$c$ plane sensitively affects the structure of the torque curves in the superconducting region.  

The effect of disorder on the magnetic torque is not observed in the normal state above approximately 5 T in both the BMDT-TTF-molecule-substituted and X-ray-irradiated samples.  
The magnetic torque curves show paramagnetic behavior on which the dHvA oscillations are superimposed and show no inclusion of magnetic impurities that might be induced by disorder.

The dHvA oscillations on the torque curves can be seen above approximately 5 T even in the vortex liquid state below the upper critical field $H_{\rm c2}$ in the clean pristine $\kappa$-($h$-BEDT-TTF)$_{2}$Cu(NCS)$_{2}$ \cite{sasaki1998}. 
From the dHvA oscillations, the cross-sectional area of the Fermi surface, the effective mass $m^{*}$, and the scattering time $\tau_{\rm dHvA}$ of the cyclotron-orbiting electrons on the Fermi surfaces are obtained from the oscillation frequency $F$, and the temperature and magnetic field dependences of the oscillation amplitude, respectively, on the basis of the Lifshitz-Kosevich formula \cite{shoenberg}.
The oscillation part of the magnetic torque $A_{\rm LK}$ excluding higher-harmonic components is described as
\begin{equation}
A_{\rm LK} \propto H^{n}R_{\rm T}R_{\rm D}R_{\rm S}\sin[2\pi(\frac{F}{H}-\gamma)],
\end{equation}
for the temperature factor $R_{\rm T} = [\lambda(m^{*}/m_{0})T/H]/\sinh[\lambda(m^{*}/m_{0})T/H]$, the Dingle factor $R_{\rm D} = \exp[-\lambda(m^{*}/m_{0})T_{\rm D}/H]$, and the spin factor $R_{\rm S} = \cos[\pi{g}(m_{\rm b}/m_{0})/2]$.  
$\gamma$ is a phase factor that is usually close to 1/2.
Here, $\lambda \equiv 2 \pi^{2} c k_{\rm B}/e \hbar =$ 14.69 T/K, $m_{0}$ and $m_{\rm b}$ are the free electron and band masses, respectively, and $T_{\rm D}$ is the Dingle temperature, related to the scattering time $\tau_{\rm dHvA}$ by the relation $T_{\rm D}= \hbar / 2\pi k_{\rm B} \tau_{\rm dHvA}$.  
For the torque-dHvA oscillation in three and two dimensions, $n$ is 3/2  and 1, respectively. 
Here, $n =$ 1 for two dimensions is used.

\begin{figure}
\includegraphics[width=1\linewidth]{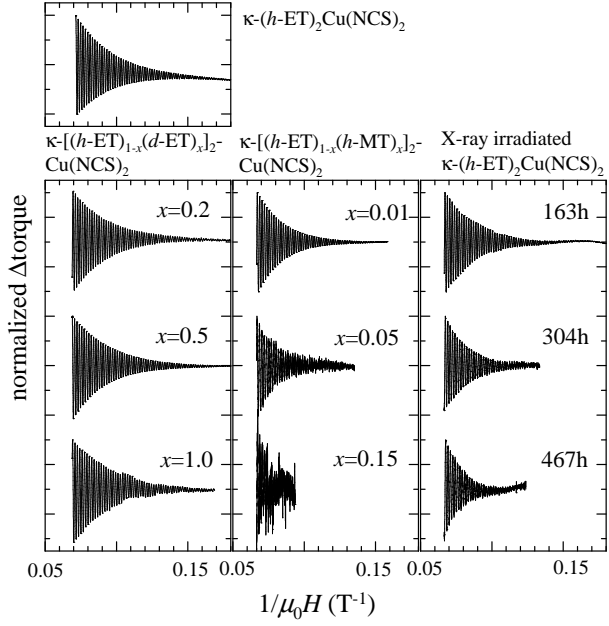}
\caption{Magnetic torque dHvA oscillations at $T =$ 0.44 K in (upper) pristine $\kappa$-($h$-BEDT-TTF)$_{2}$Cu(NCS)$_{2}$, (left) $\kappa$-[($h$-BEDT-TTF)$_{1-x}$($d$-BEDT-TTF)$_{x}$]$_{2}$Cu(NCS)$_{2}$, (center) $\kappa$-[($h$-BEDT-TTF)$_{1-x}$($h$-BMDT-TTF)$_{x}$]$_{2}$Cu(NCS)$_{2}$, and (right) $\kappa$-($h$-BEDT-TTF)$_{2}$Cu(NCS)$_{2}$ irradiated with X-rays.  The magnetic fields are applied perpendicular to the $b$-$c$ plane. The amplitude of the oscillations is normalized by the value of each oscillation at 14 T.}
\end{figure}

Figure 4 shows the dHvA oscillations of pristine $\kappa$-($h$-BEDT-TTF)$_{2}$Cu(NCS)$_{2}$ in the upper panel, $\kappa$-[($h$-BEDT-TTF)$_{1-x}$($d$-BEDT-TTF)$_{x}$]$_{2}$Cu(NCS)$_{2}$ in the left panel, $\kappa$-[($h$-BEDT-TTF)$_{1-x}$($h$-BMDT-TTF)$_{x}$]$_{2}$Cu(NCS)$_{2}$ in the center panel, and $\kappa$-($h$-BEDT-TTF)$_{2}$Cu(NCS)$_{2}$ irradiated with X-rays in the right panel.  
The results of the oscillation frequency $F =$ 602 $\pm$ 2 T, the effective mass $m^{*} =$ (3.2 $\pm$ 0.2)$m_{0}$, and the Dingle temperature $T_{\rm D} =$  0.48 $\pm$ 0.04 K, corresponding to the scattering time $\tau_{\rm dHvA} =$ 2.5 $\pm$ 0.2 ps in the pristine sample, reproduce well the results in previous reports \cite{oshima,sasaki1998,sasaki1990,caulfield}. 
In the present magnetic field region below 15 T, we observe oscillations with a single frequency that have been attributed to the $\alpha$-orbit of the cylindrical Fermi surface.  
No magnetic breakdown orbit $\beta$, which has been observed at higher magnetic fields, is detected because of the lower experimental magnetic fields than the magnetic breakdown field \cite{sasaki1990,sasaki1991,caulfield,meyer,harrison}.  

From the observed dHvA oscillations shown in Fig. 4, the oscillation frequency, effective mass, and scattering time are obtained on the basis of the Lifshitz-Kosevich formula.  
These parameters are summarized in Fig. 5, which are plotted in comparison with the substitution content $x$ and irradiation time.

\begin{figure}
\includegraphics[width=1\linewidth]{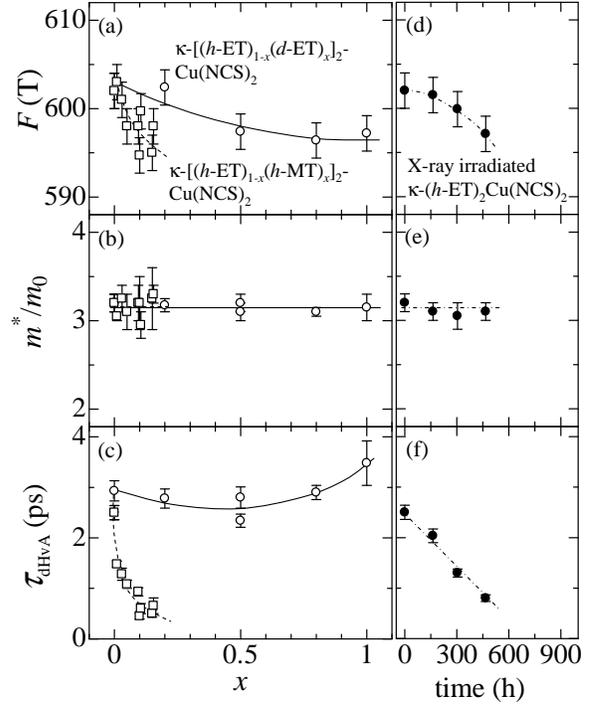}
\caption{Molecule substitution and X-ray irradiation time dependences of dHvA frequency $F$, effective mass ratio $m^{*}/m_{0}$, and scattering time ${\tau}_{\rm dHvA}$ obtained by dHvA measurements in (a), (b), (c) $\kappa$-[($h$-BEDT-TTF)$_{1-x}$($d$-BEDT-TTF)$_{x}$]$_{2}$Cu(NCS)$_{2}$ (open circles) and $\kappa$-[($h$-BEDT-TTF)$_{1-x}$($h$-BMDT-TTF)$_{x}$]$_{2}$Cu(NCS)$_{2}$ (open squares) and in (d), (e), (f) $\kappa$-($h$-BEDT-TTF)$_{2}$Cu(NCS)$_{2}$ irradiated with X-rays (filled circles), respectively. The curves are guides for the eyes. }
\end{figure}

\subsubsection{Substitution effect with deuterated BEDT-TTF molecule, $\kappa$-[($h$-BEDT-TTF)$_{1-x}$($d$-BEDT-TTF)$_{x}$]$_{2}$Cu(NCS)$_{2}$}

The oscillation frequency shows a small monotonic decrease with $x$ from $F =$ 602 $\pm$ 2 T for $x =$ 0 to  597 $\pm$ 2 T for $x =$ 1, which is consistent with a previous report \cite{biggs}.
This monotonic change in $F$ with $x$ can be explained by the chemical pressure effect.  
The sample with $x =1$ is considered to have a narrower bandwidth than the sample with $x =$ 0.  
This is because the full deuteration of the BEDT-TTF molecule in $\kappa$-(BEDT-TTF)$_{2}$$X$ induces a negative pressure effect of approximately 15-20 MPa \cite{sasaki2009stam}. 
The change in $F$ with pressure in $\kappa$-($h$-BEDT-TTF)$_{2}$Cu(NCS)$_{2}$ has been investigated, and the pressure causes $F$ to increase by approximately 0.3\%/10 MPa \cite{caulfield}.  
This pressure effect owing to the full deuteration from $x =$ 0 to 1 is expected to lead to a decrease of approximately 3 - 4 T at approximately $F \simeq$ 600 T. 
The observed change in $F$ is consistent with this estimation within the accuracy of the measurements of $F$.  

The effective mass $m^{*}$ of approximately 3.2$m_{0}$ does not depend on $x$, as shown in Fig. 5(b), whereas the scattering time $\tau_{\rm dHvA}$ shows a concave variation with $x$ even though a large sample dependence exists at $x =$ 0 and 0.5.  
For this deuterated molecule substitution, we do not expect a very large disorder effect on the electronic properties, as mentioned for superconductivity in the previous section.  
The randomness induced by the substitution, however, may increase the electron scattering to some extent.  
In such a case, the largest disorder effect may occur at approximately $x =$ 0.5. 
Actually, the concave feature of $\tau_{\rm dHvA}$ with $x$ indicates the possibility of the phenomenological relation $1/\tau \propto x(1-x)$, known as the Nordheim relation \cite{nordheim}, in the residual resistivity of metal alloys such as copper-gold \cite{elio,swihart}. 
The deuterated molecule substitution, however, does not cause a large spatial modulation by the local impurity potential, which results in large-angle scattering $\tau_{\rm imp}$ for the resistivity and superconductivity. 
Meanwhile, the scattering time obtained by the dHvA effect should also be affected by the small-angle scattering in addition to the large-angle scattering.
The observed concave feature of $\tau_{\rm dHvA}$ with $x$ may be explained by the phase shift of the oscillations discussed for the weak inhomogeneity in the metal alloy system \cite{shoenberg,watts}.  

Recently, an alternative model for the decrease in scattering time obtained from dHvA oscillations has been proposed on the basis of the inhomogeneous electronic state \cite{harrison2001,singleton2001}. 
This model shows that the decrease in dHvA amplitude is caused by the interference of dHvA oscillations with slightly different frequencies, which originate from the statistical distribution of the electronic inhomogeneity in the bulk sample.  
The present observation may be an example of such a case because the change in $F$ with $x$ is likely to cause microscopic inhomogeneity in the electronic states.  
Neither macroscopic inhomogeneity nor a large disorder effect, however, is expected because the substitution does not produce any inhomogeneous feature in $T_{\rm c}$ and ${\Delta}T_{\rm c}$, as shown in Fig. 2.

\subsubsection{Substitution effect with $h$-BMDT-TTF molecule, $\kappa$-[($h$-BEDT-TTF)$_{1-x}$($h$-BMDT-TTF)$_{x}$]$_{2}$Cu(NCS)$_{2}$}

The $h$-BMDT-TTF molecule substitution has a much stronger influence on the dHvA oscillations than the $d$-BEDT-TTF molecule substitution.  
The BMDT-TTF molecule substitution markedly reduces the oscillation amplitude in comparison with the pristine $\kappa$-($h$-BEDT-TTF)$_{2}$Cu(NCS)$_{2}$, and then the magnetic fields in which the oscillations become observable shift to higher magnetic fields.  

As is clearly seen in Fig. 5(a), the oscillation frequency decreases approximately by 1\% with increasing substitution ratio $x$ up to 0.15.   
Some possible causes are considered for the decrease in $F$.
First, the negative chemical pressure induced by the BMDT-TTF molecule substitution may be expected to increase the lattice parameters, similar to the case of the $d$-BEDT-TTF molecule substitution.
The lattice parameters of the BMDT-TTF-molecule-substituted sample, however, show no significant difference from those of the pristine sample.\cite{MT-lattice}
In addition, a reduction in unit cell volume by the substitution is expected because the BMDT-TTF molecule is smaller than the BEDT-TTF molecule.  
Actually the crystal $\kappa$-($h$-BMDT-TTF)$_{2}$Cu[N(CN)$_{2}$]Br has a smaller unit cell volume \cite{BMDT} than $\kappa$-($h$-BEDT-TTF)$_{2}$Cu[N(CN)$_{2}$]Br, which is the same $\kappa$-type organic superconductor \cite{Br} as the present material.  
In this case, a positive chemical pressure is expected for the smaller-molecule substitution, that is, $F$ should increase with increasing $x$. 
This tendency is in contrast to the observation.  

The other possible origin of the decrease in $F$ is the slight modifications of the Brillouin zone and the anisotropy of the transfer energies.  
The slight modifications of the Brillouin zone and the anisotropy of the transfer energies sensitively lead to a variation in the cross-sectional area of the $\alpha$-orbit because the $\alpha$-orbit is formed in the extended Brillouin zone \cite{oshima}. 
In the sister compound system mentioned above, the crystal lattice of $\kappa$-($h$-BMDT-TTF)$_{2}$Cu[N(CN)$_{2}$]Br is monoclinic \cite{BMDT}, while that of $\kappa$-($h$-BEDT-TTF)$_{2}$Cu[N(CN)$_{2}$]Br is orthorhombic \cite{Br}. 
Considering this difference in the crystal lattice, the substitution of the smaller BMDT-TTF molecule in the present material might not cause a simple isotropic contraction of the lattice parameters. 

An alternative idea may be an inhomogeneous modulation of local electronic states around the substitution site.  
If the substituted BMDT-TTF molecules do not provide the normal amount of electron charges, +0.5$e$/donor molecule, to anion molecules, the hole charge on the BEDT-TTF molecules around the substitution site of the BMDT-TTF molecule also varies to maintain the formal charge transfer value in the crystal. 
Such local modulation of the electronic states may cause a small change in the Fermi surface.
At the moment, we do not have a conclusive explanation for the small change in $F$ with $x$.

In contrast to the deuterated BEDT-TTF molecule substitution, the scattering time decreases very rapidly from $\tau_{\rm dHvA} \simeq$ 2.5 ps for $x =$ 0 to 0.5 ps for $x =$ 0.15.  
The decrease in $\tau_{\rm dHvA}$ is considered to be mostly owing to the impurity scattering induced by the substituted BMDT-TTF molecules, which may act as point scattering centers.  
With increasing amount of the substitution, however, an inhomogeneous distribution of the substitution may occur in addition to the impurity scattering.  
The inhomogeneity could influence the decrease in $\tau_{\rm dHvA}$ in addition to the impurity scattering.  
This additional damping effect may be the case of the inhomogeneity in space discussed above \cite{harrison2001,singleton2001}.
As we mentioned in the previous section for superconductivity, the increase in ${\Delta}T_{\rm c}$  shown in Fig. 2 may support the presence of inhomogeneity. 

Here we estimate the mean free path $l$ because we could observe dHvA oscillations even though a large disorder from impurities and inhomogeneity are induced in the crystal.  
The mean free path $l$ is calculated to be $\simeq$ 25 nm for $x =$ 0.15, where $l \equiv v_{\rm F}\tau_{\rm dHvA}$ and $v_{\rm F} \simeq 5 \times 10^{4}$ m/s for the $\alpha$-orbit.  
$l$ is still larger than the in-plane coherence length $\xi \simeq$ 5 -- 6 nm \cite{lang,yone3}.
This condition of $l \gg \xi$ indicates that the superconductivity of the sample with the BMDT-TTF substitution should be in a clean limit.  

\subsubsection{X-ray irradiation effect in $\kappa$-($h$-BEDT-TTF)$_{2}$Cu(NCS)$_{2}$}

The X-ray irradiation may cause molecular defects in the crystal \cite{mihaly}. 
The disorder effect of the defects suppresses superconductivity, as shown in the previous section.  
The dHvA oscillations are certainly affected by disorder.  
The oscillation frequency $F$ decreases by approximately 1\% after 467 h of X-ray irradiation.  
This change in $F$ may originate from the local imbalance of the charge transfer between the BEDT-TTF donor and Cu(NCS)$_{2}$ anion layers.  
It has been expected that the molecular defects will be introduced mainly into the anion molecules related to the CN ligand for Cu \cite{yone2009,yone2010}.  
In the case of the defects in anion molecules, less charge transfer from BEDT-TTF than the normal value of +0.5$e$/molecule may arise locally in the crystal owing to a space modulation of the mean anionicity of polymeric anions.  
This local imbalance of the charge transfer may induce effective carrier doping \cite{sasaki_Cl_x-ray} and potential modulation in space as disorder \cite{sasaki_Br_x-ray} in the conductive BEDT-TTF layers.
It is difficult in principle to evaluate the number of defects and the expected change in the charge transfer value, which are induced by X-ray irradiation.  
In comparison to the molecule substitution with $h$-BMDT-TTF, however, disorder caused by approximately 500 h of X-ray irradiation seems to roughly correspond to that with the substitution content $x =$ 0.05 -- 0.1 in terms of the change in $F$.  
This quantitative relation has also been similarly found in the behaviors of $T_{\rm c}$, and ${\Delta}T_{\rm c}$ as mentioned in the previous section.  

The effective mass does not change with the X-ray irradiation in the same way as the $h$-BMDT-TTF molecule substitution.  
On the other hand, a large decrease in $\tau_{\rm dHvA}$ is observed.  
This decrease in $\tau_{\rm dHvA}$ could also be explained by the disorder effect caused by the X-ray irradiation, which has been discussed as a local modulation of the charge transfer by defects in anion molecules.  
In fact, the reduction in $\tau_{\rm dHvA}$ is consistent with the changes in $T_{\rm c}$ and $F$ for the $h$-BMDT-TTF molecule substitution discussed above.  

To understand the effect of the X-ray irradiation on the electronic states in detail, further investigation of the morphology and nature of molecular defects is necessary in the future.

\subsection{Relation between superconductivity and scattering time}

In this section, we discuss the relation between superconductivity and the scattering time $\tau_{\rm dHvA}$ examined in the molecule-substituted and X-ray-irradiated $\kappa$-(BEDT-TTF)$_{2}$Cu(NCS)$_{2}$.  
We emphasize here that the disorder level in the present crystals is not too large to cause a direct influence on superconductivity through the modification of the electronic states at the Fermi level. \cite{Tc_FS}
This is because the clean limit condition of the superconductivity can be applied without significant changes in $F$ and $m^{*}$ on the Fermi surface.  
Note in particular that the crystals in which the disorders are introduced are still clean for quasi-particles since dHvA oscillations are observable.  
Therefore, we assume in this section that most of the decrease in $T_{\rm c}$ is affected markedly by changing the quasi-particle scattering, at least at the low disorder level with $\tau_{\rm dHvA} >$ 1 ps.

In the case of the deuterated BEDT-TTF molecule substitution, almost no disorder effect is observed on the superconductivity.
The partial substitution acts as a parameter for changing the bandwidth, which becomes narrower with the substitution.  
In other words, this result microscopically demonstrates that the deuterated BEDT-TTF molecule substitution can be regarded as a negative chemical pressure applied to the crystal with minimal effect on disorder. 
This fact is also important for discussions on the electronic phase separation in $\kappa$-(BEDT-TTF)$_{2}$Cu[N(CN)$_{2}$]Br as has been observed near the Mott transition \cite{sasaki2004PRL,sasaki2005JPSJ,sasaki2009stam}, in which the bandwidth is controlled by the deuterated molecule substitution.  
This is because the present results ensure that the phase separation in space takes place not by the structural inhomogeneity but by a purely electronic origin in the first-order metal-Mott insulator transition.

Next, we discuss the suppression of superconductivity by the $h$-BMDT-TTF molecule substitution and X-ray irradiation from the viewpoint of nonmagnetic impurity scattering.  
The suppression of superconductivity by disorder in organic superconductors has been investigated in several materials using different means of introducing disorder \cite{powell}.  
In most cases, however, the amount of disorder has been evaluated from the residual resistivity.  
In the case of the superconductor, one needs to apply a magnetic field to suppress the superconductivity or to assume the extrapolation function of the temperature dependence of the resistivity curve at higher temperatures above $T_{\rm c}$ to obtain the residual resistivity at sufficiently low temperatures.  
Moreover, from the intralayer resistivity in this class of layered organic superconductors, it is difficult to obtain a reliable value owing to the experimental difficulties in ensuring a homogeneous current distribution. 
For the interlayer resistivity mentioned above, we need to know the interlayer transfer integral $t_{\perp}$, which should be obtained independently from other experiments.  
Moreover, it is usually difficult to determine the value accurately.  
In this situation, one important advance made by the present study is that $\tau_{\rm dHvA}$ can be used to evaluate the suppression of superconductivity.
To the best of the authors' knowledge, this is the first approach to evaluating $T_{\rm c}$ using $\tau_{\rm dHvA}$ for not only organic superconductors but also inorganic superconductors.

Anderson's theorem indicates that nonmagnetic impurities do not change $T_{\rm c}$ in $s$-wave conventional superconductors \cite{anderson}. 
On the other hand, magnetic impurities reduce $T_{\rm c}$ because they break the time-reversal symmetry of Cooper pairs \cite{maple}. 
The quantitative behavior of $T_{\rm c}$ is given by the AG formula, which was originally used to describe the magnetic impurity effect on the reduction in $T_{\rm c}$ in conventional superconductors \cite{AG,tinkham}.  
In an unconventional case with non $s$-wave symmetry, however, the same formulation has been expected to be applicable to the nonmagnetic impurity effect on the reduction in $T_{\rm c}$  \cite{maple,millis,radtke,powell}.
Indeed, a noticeable reduction in $T_{\rm c}$ by nonmagnetic impurities has been successfully applied to the AG formula to confirm the unconventional superconductivity \cite{mackenzie}. 
It is therefore interesting to examine whether the present substitution and X-ray irradiation effects are also occur.  
The transition temperature $T_{\rm c}^{\rm AG}$ affected by the impurity scattering is described by the AG formula as

\begin{equation} 
\ln \left(\frac{T_{\rm c0}}{T_{\rm c} ^{\rm AG}} \right) = \psi \left(\frac{1}{2}+\frac{\hbar}{4\pi k_{\rm B} T_{\rm c} ^{\rm AG}} \cdot \frac{1}{\tau_{\rm AG}} \right) - \psi \left(\frac{1}{2} \right), 
\end{equation}
where $T_{\rm c0}$ is $T_{\rm c}$ in an ideally pure system with $\tau_{\rm AG} \rightarrow \infty$ and $\psi (x)$ is the digamma function. 
In the AG formula, $\tau_{\rm AG}$ is supposed to be the large-angle impurity scattering in principle. 
In short, $\tau_{\rm AG} \simeq \tau_{\rm imp}$. 

The scattering time for electrons in metals includes multiple contributions from different scattering processes \cite{powell,chang}.  
Each physical quantity obtained by different experiments has a characteristic contribution from different scattering processes to the evaluated scattering time.
The residual resistivity, for example, is dominated mainly by the elastic large-angle scattering by impurities.  
On the other hand, the scattering time obtained from the dHvA oscillations includes additional contributions from smaller-angle scattering by dislocations, weak potential modulation in space, and so on.  
Therefore, the scattering rate $1/\tau_{\rm dHvA}$ can be described on the basis of Matthiessen's rule as
\begin{equation}
1/\tau_{\rm dHvA} = 1/\tau_{\rm imp} + 1/\tau_{\rm other},
\end{equation}
where $\tau_{\rm other}$ denotes the scattering time excluding the impurity scattering.

\begin{figure}
\includegraphics[width=1\linewidth]{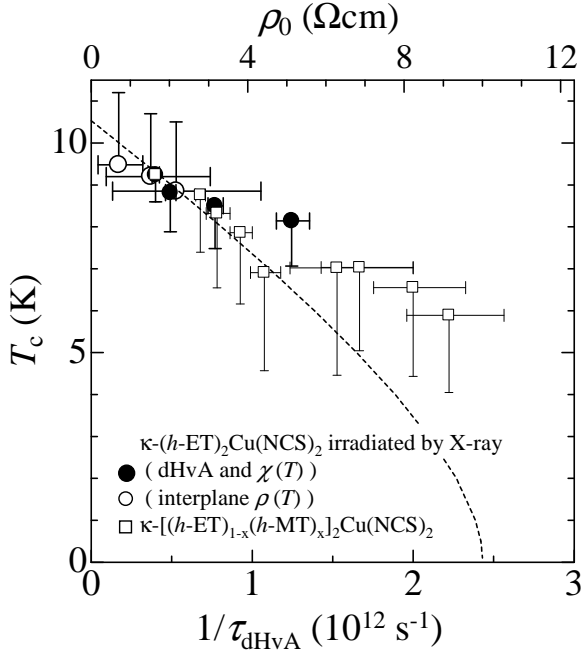}
\caption{Relation between $T_{\rm c}$ and $1/{\tau}_{\rm dHvA}$ for samples of $\kappa$-[($h$-BEDT-TTF)$_{1-x}$($h$-BMDT-TTF)$_{x}$]$_{2}$Cu(NCS)$_{2}$ (open squares) and $\kappa$-($h$-BEDT-TTF)$_{2}$Cu(NCS)$_{2}$ irradiated with X-rays (filled circles). The vertical bars indicate the half-width of the transition curves of the magnetic susceptibility as shown in Fig. 1(a). The dashed curve is calculated from the AG formula for $\tau_{\rm AG} = \tau_{\rm dHvA}$ and $T_{\rm c}^{\rm AG} =$ 10.5 K. The results obtained from the interlayer resistivity, shown in Fig. 1(c) (open circles), are plotted in relation to the interlayer residual resistivity $\rho_{0}$ on the upper axis. The vertical bars for the resistivity data indicate the onset and offset temperatures of the resistive transition curves. The relation between $\rho_{0}$ and $\tau_{\rm dHvA}$ is described in the text.}
\end{figure}

Figure 6 shows the relation between the scattering time $\tau_{\rm dHvA}$ and $T_{\rm c}$ in the samples substituted with $h$-BMDT-TTF (open squares) and irradiated with X-rays (filled circles).  
$T_{\rm c}$ is determined by magnetic susceptibility measurements and the vertical bars indicate the half-width of the transition curves as shown in Figs. 1(a) and 1(b). 
The dashed curve is the calculated $T_{\rm c}^{\rm AG}$ based on the AG formula of eq. (2) with $T_{\rm c0} =$ 10.5 K, where $\tau_{\rm dHvA}$ is applied for $\tau_{\rm AG}$ in the AG equation. 
In the same figure, we plot the suppression of $T_{\rm c}$ by the X-ray irradiation, which is obtained from the interlayer resistivity in comparison with the residual resistivity $\rho_{0}$ as shown in Fig. 1(c).  
The residual resistivity at $T = 0$ is evaluated from the quadratic extrapolation of the resistivity curve as $\rho(T) = \rho_{0} + AT^{2}$ below approximately 20 K.
The vertical bars for the resistivity data show the onset and offset temperatures of the resistive transition curves. 
The $T_{\rm c}$ values evaluated in the previous report by Analytis {\it et al.} \cite{analytis} correspond to the middle temperature of the bars.
The correspondence between $\rho_{0}$ (upper transverse axis) and $1/\tau_{\rm dHvA}$ (lower axis) is considered on the basis of the previous arguments on the interlayer resistivity \cite{powell,mckenzie1998}.
The residual resistivity for interlayer transport is given by 
\begin{equation}
\rho_{0} = \frac{\pi\hbar^{4}}{2e^{2}m^{*}dt_{\perp}^{2}}\frac{1}{\tau_{\rm imp}},
\end{equation}
where $d$ is the interlayer spacing and $t_{\perp}$ is the interlayer transfer integral.  
By following eq. (4), the linear relation between $\rho_{0}$ and $\tau_{\rm imp}$, which corresponds to $\tau_{\rm dHvA}$ in Fig. 6, is given by the parameter $t_{\perp} =$ 0.048 meV (0.034 meV) while taking $d$ = 1.522 nm \cite{watanabe} and $m^{*} = 3.2m_{0}$ ($6.5m_{0}$).  
Here, $m^{*} = 6.5m_{0}$ in the case for the magnetic breakdown orbit at higher magnetic fields \cite{sasaki1990,caulfield,meyer,harrison}.  
Although it is a rough estimation, the applied $t_{\perp}$ is in reasonably good agreement with the value ($0.03 \pm 0.01$ meV) reported by Analytis {\it et al}. \cite{analytis} and also with an independent determination of $t_{\perp} \simeq$ 0.04 meV by an angular-dependent magnetotransport experiment \cite{singleton2002,goddard}. 
A small difference in the correspondence between $\rho_{0}$ and $1/\tau_{\rm dHvA}$ in the X-ray-irradiated samples may be caused by the contribution of $1/\tau_{\rm other}$ in eq. (3).
This is because $\rho_{\rm 0}$ tends to be smaller as $\rho_{0} \propto (1/\tau_{\rm dHvA} - 1/\tau_{\rm other})$ in practice.
For such quantitative arguments on the transport properties, we need further investigation in the future while considering the pristine sample-to-sample dependence.

The suppression of $T_{\rm c}$ by both molecule substitution and X-ray irradiation is similarly observed for $\tau_{\rm dHvA}$.  
The dependence of $T_{\rm c}$ on $1/\tau_{\rm dHvA}$ is almost linear in the initial region below $1/\tau_{\rm dHvA} \simeq$ 1 $\times$ 10$^{12}$ s$^{-1}$.
This behavior is consistent with the AG formula of eq. (2).
The previous results \cite{analytis} for $\rho_{0}$ are also consistent with the present results for $\tau_{\rm dHvA}$ if $\tau_{\rm imp} = \tau_{\rm dHvA}$ is assumed. 
These observations indicate that, at least in the small-disorder region, the nonmagnetic impurity scattering induced by molecule substitution and X-ray irradiation works as the pair-breaker expected in the unconventional superconductivity.  
In addition, the good correspondence among the scattering time $\tau_{\rm imp}$, $\tau_{\rm dHvA}$, and $\tau_{\rm AG}$ suggests that the contribution of $1/\tau_{\rm other}$ to $1/\tau_{\rm dHvA}$ in eq. (3) is relatively small and thus $\tau_{\rm dHvA} \simeq \tau_{\rm imp}$ holds in a weak-disorder region below $1/\tau_{\rm dHvA} \simeq$ 1 $\times$ 10$^{12}$ s$^{-1}$.  

Note, however, that both the present results and the previous results by Analytis {\it et al.} \cite{analytis} deviate from the AG formula in the larger-disorder region above $1/\tau_{\rm dHvA} \simeq$ 1 $\times$ 10$^{12}$ s$^{-1}$, in which the decrease in $T_{\rm c}$ becomes smaller with increasing $1/\tau_{\rm dHvA}$ and $\rho_{0}$. 
In addition, such deviation seems to start at a different $1/\tau_{\rm dHvA}$ in the X-ray-irradiated and BMDT-TTF-molecule-substituted samples. 
Looking more closely at the cases of the X-ray irradiation, the deviation observed in the present and previous studies \cite{analytis} starts to appear at somewhat different values of $1/\tau_{\rm dHvA}$ and at equivalent $\rho_{0}$ values.  
In their previous study, Analytis {\it et al.} \cite{analytis} proposed a possible mixed order parameter with both non-$s$-wave unconventional and $s$-wave components for the deviation from the AG formula.  
In addition to such an explanation, we note other possible causes of the deviation from the AG formula as follows.  
These are based on the experimental difficulty in the accurate evaluation of the scattering time from either the residual resistivity or quantum oscillation measurement.
First, for the deviation from the AG formula, the insensitive behavior of $T_{\rm c}$ at larger $1/\tau_{\rm dHvA}$ may be caused by the increasing contribution of $1/\tau_{\rm other}$ to $1/\tau_{\rm dHvA}$ as indicated in eq. (3), in comparison with $1/\tau_{\rm imp} \simeq 1/\tau_{\rm AG}$.  
The additional contribution $\tau_{\rm other}$ may result from the effect of the inhomogeneous potential modulation in bulk samples discussed above, which induces a damping of the dHvA oscillations but does not markedly affect the reduction in $T_{\rm c}$.  
Therefore, the different effect of $\tau_{\rm other}$ on the dHvA oscillations and superconductivity may be the origin of the observed weak $1/\tau_{\rm dHvA}$ dependence at larger disorder levels.  

Second, we comment on the relevance of $\rho_{0}$ to the evaluation of the scattering time using the AG formula in the X-ray irradiation experiments by Analytis {\it et al.}\cite{analytis}
Recently, Sano {\it et al.} have reported that the disorder introduced by the X-ray irradiation of the organic superconductor $\kappa$-(BEDT-TTF)$_{2}$Cu[N(CN)$_{2}$]Br enhanced the electron localization, and then the superconductor became an Anderson-type insulator with increasing irradiation dose. \cite{sasaki_Br_x-ray}
Such a localization effect is enhanced toward the Mott critical point of the insulator-metal transition with increasing electron correlations. 
Considering the localization effect, the residual resistivity of the strongly correlated electron system with disorder does not relate simply to the impurity scattering.  
In the present case, the localization effect caused by introducing disorder in $\kappa$-(BEDT-TTF)$_{2}$Cu(NCS)$_{2}$ might be weaker than that in $\kappa$-(BEDT-TTF)$_{2}$Cu[N(CN)$_{2}$]Br because the former has weaker electron correlations owing to its broader bandwidth in comparison with the narrower one in the latter. \cite{sasaki_Br_x-ray}
Although the weaker effect of the localization is expected in the present material, the residual resistivity may become larger than the value derived by the impurity scattering, as described in eq. (4) with increasing level of disorder. 
This additional increase in $\rho_{0}$ may be one of the reasons for the deviation of $T_{\rm c}$ from the AG formula.

Finally, we mention the reduction in $T_{\rm c}$ in comparison with that in the case of metal superconductors with magnetic transition-metal impurities.  
A large suppression of $T_{\rm c}$ by a small amount of magnetic impurities is known to occur in Al, Zn, and other metal superconductors.\cite{boato}
For example, the rate of reduction of $T_{\rm c}$ amounts to approximately $-3$ K/0.01 at.\% Mn in Zn.\cite{boato}  
In such a case, the rate of reduction can be expected to be proportional to $J^2S_{\rm e}S_{\rm i}/E_{\rm F}$, where $J$ is the exchange coupling between the impurity and conduction electron spins, and $S_{\rm e}$ and $S_{\rm i}$ are the spins of the conduction electrons and impurities, respectively.\cite{boato,deGennes} 
The value of $J \sim$ 1 eV in Zn-Mn is approximately two orders of magnitude larger than that in organic conductors ($J \sim 0.004 - 0.02$ eV)\cite{powell}.
Therefore, such a magnetic impurity effect on $T_{\rm c}$ may not account for the present large reduction in $T_{\rm c}$ in organic superconductors even if small impurity spins are induced in molecular defects.
  
\section{Conclusion}
The suppression of superconductivity is investigated in the organic superconductor $\kappa$-(BEDT-TTF)$_2$Cu(NCS)$_2$, in which different types of disorder are introduced by molecule substitution and X-ray irradiation.  
To determine the relation between superconductivity and disorder, we systematically examined the electronic states of $\kappa$-(BEDT-TTF)$_2$Cu(NCS)$_2$ with introducing disorder.
The disorder induces a random electronic potential modulation in space with minimal effect on the electronic states at the Fermi level.  
A large reduction in $T_{\rm c}$ with a linear dependence on $1/\tau_{\rm dHvA}$ evaluated using the dHvA effect is found in the small-disorder region below $1/\tau_{\rm dHvA} \simeq$ 1 $\times$ 10$^{12}$ s$^{-1}$ in both the BMDT-TTF molecule-substituted and X-ray-irradiated samples. 
The observed linear relation between $T_{\rm c}$ and $1/\tau_{\rm dHvA}$ is in agreement with the AG formula.  
This observation is reasonably consistent with the unconventional superconductivity of this organic superconductor that has been suggested by theoretical and experimental investigations so far.  

However, a deviation from the AG formula is observed in the large-disorder region above $1/\tau_{\rm dHvA} \simeq$ 1 $\times$ 10$^{12}$ s$^{-1}$, which reproduces the results of a previous transport study. \cite{analytis}
As reported by Analytis {\it et al.} \cite{analytis}, this deviation suggests that superconductivity does not simply obey the AG formula on the basis of the single non-$s$-wave order parameter, but involves a mixed parameter of $s$- and non-$s$-waves.  
Furthermore, in consideration of the reduction in $T_{\rm c}$ with an increase in the scattering time, there are experimental difficulties in the accurate evaluation of the scattering time, thereby affecting the superconductivity inherently. 
The present dHvA study was motivated by the quantitative evaluation of the scattering time, which could be obtained more reliably without the complex assumptions needed in the transport study.  
Some of the difficulties and uncertainties in the evaluation of the scattering time might be related to the deviation from the AG formula. 
It is clear that the superconductivity of this class of organic materials should be investigated further to completely understand all the observed phenomena in the past 25 years. 

\section*{Acknowledgments}

The authors thank M. Lang, J. M{\"u}ller, T. Nakamura, and K. Yakushi for valuable discussions.
Part of this work was performed at the High Field Laboratory for Superconducting Materials (HFLSM), IMR, Tohoku University.  
This work was partly supported by Grants in-Aid for Scientific Research (Nos. 20340085 and 21110504) from JSPS and MEXT, Japan.


\begin{thebibliography}{99}
\bibitem{chi} In the case of the X-ray-irradiated sample shown in Fig. 1(d), the demagnetization factor may change slightly, probably owing to the rounded crystal edge in the course of the repeat experiments of the X-ray irradiation at room temperature and the magnetization measurements at low temperatures for the same crystal.
\bibitem{Tc_FS} Hydrostatic pressures applied to the pristine sample lower $T_{\rm c}$ by approximately 4 K/kbar concurrently with a 3\% increase in $F$ and a 14\% decrease in $m^{*}$ for the $\alpha$-orbit.\cite{caulfield,biggs} It seems reasonable to suppose here that the nature of the $\gamma$-orbit, which is another orbit of the Fermi surface, does not change so much, similarly to the $\alpha$-orbit, although the $\gamma$-orbit is not observed as the magnetic breakdown $\beta$-orbit.  This is because both the $\gamma$- and $\alpha$-orbits originate from the same degenerated band.  No change in $m^{*}$ and a small decrease in $F$ in the $\alpha$-orbit of the present samples with the introduction of disorder indicate that we can reasonably exclude the possibility of significant modulation of the electronic states at the Fermi level as the origin of the large decrease in $T_{\rm c}$.
\end{thebibliography}

\end{document}